    \documentclass[twocolumn]{aastex7}

\submitjournal{AJ}

\shorttitle{CRF-K-2025 Catalog }
\shortauthors{Gordon et al.}

\begin{document}

 \title{The Celestial Reference Frame at K Band: The CRF-K-2025 Catalog }

\correspondingauthor{David Gordon}

\author[0000-0001-8009-995X]{David Gordon}
\affiliation{U.S. Naval Observatory}
\email{david.gordon126.civ@us.navy.mil}

\author[0000-0001-9885-4220]{Aletha de Witt }
\affiliation{South African Radio Astronomy Observatory (SARAO) }
\email{aletha.dewitt@dsti.gov.za}

\author[0000-0003-0828-1401]{Christopher S. Jacobs}
\affiliation{Jet Propulsion Laboratory, California Institute of Technology/NASA }
\email{christopher.s.jacobs@jpl.nasa.gov  }

\author[0000-0001-8660-649X]{Hana Kr\'{a}sn\'{a}}
\affiliation{TU Wien, Department of Geodesy and Geoinformation }
\email{Hana.Krasna@geo.tuwien.ac.at}

\author{Alessandra Bertarini}
\affiliation{German Space Agency within DLR }
\email{Alessandra.Roy@dlr.de }

\author[0000-0002-0233-6937]{Jamie McCallum }
\affiliation{University of Tasmania }
\email{jamie.mccallum@utas.edu.au}

\author{Jonathan Quick }
\affiliation{South African Radio Astronomy Observatory (SARAO) }
\email{jon@hartrao.ac.za}


\author[0000-0003-1136-5016]{Cristina Garc\'{i}a-Mir\'{o}}
\affiliation{Observatorio Astron\'{o}mico Nacional, IGN, Spain}
\email{c.garciamiro@oan.es}

\author{Taehyun Jung}
\affiliation{Korea Astronomy and Space Science Institute, Korea}
\email{thjung@kasi.re.kr}

\author[0000-0001-6094-9291]{Jeffrey A. Hodgson}
\affiliation{Sejong University, Korea }
\email{jhodgo@gmail.com}

\author[0009-0002-1871-5824]{Whee Yeon Cheong}
\affiliation{Korea Astronomy and Space Science Institute, Korea}
\email{wheeyeon@kasi.re.kr}

\author[0000-0002-6269-594X]{Sang-Sung Lee}
\affiliation{Korea Astronomy and Space Science Institute, Korea}
\email{sslee@kasi.re.kr}

\author[0000-0003-1157-4109]{Do-Young Byun}
\affiliation{Korea Astronomy and Space Science Institute, Korea}
\email{bdy@kasi.re.kr}



\begin{abstract}
We present an updated K band (24 GHz) celestial reference frame (CRF) constructed
from $\sim$3.5 million Very Long Baseline Interferometry (VLBI) observations 
collected during 211 observing epochs between May 2002 and December 2025 using
the Very Long Baseline Array (VLBA), the HARTRAO-HOBART26 baseline, 
the HARTRAO-YEBES40M baseline, and the Korean VLBI Network (KVN) augmented
with several other VLBI stations.
We have successfully observed and determined precise angular coordinates 
for 1317 compact extragalactic radio sources, essentially 
quasars, covering the full sky.
This updated K band catalog is designated as CRF-K-2025.
The precision of CRF-K-2025 is characterized by median scaled
uncertainties of 60 and 104 $\mu$-arc-seconds ($\mu$as) in right ascension 
and declination, respectively. 
The increase in number of observations and sensitivity over earlier 
K band campaigns has resulted in a catalog with 493 additional sources
and a precision approximately 25\% better than the ICRF3-K catalog, and 
similar to the ICRF3-SX catalog.
 At K band, these quasar radio sources generally show less extended emission
than at lower frequencies and thus can potentially
provide a more stable long term celestial reference frame than at 
the standard S/X (2.3/8.4 GHz) observing bands of ICRF3-SX.

\end{abstract}



\section{Introduction} \label{sec:intro}

High precision Very Long Baseline Interferometry (VLBI) measurements of positions of 
compact extragalactic radio sources (quasars) define and maintain the
current International Celestial Reference Frame, ICRF3 \citep{charlot2020} 
in the radio region of the electromagnetic spectrum. ICRF3 was 
approved by the International Astronomical Union in August 2018
and became the IAU's official radio reference frame in January 2019, replacing 
ICRF2 \citep{fey2009,fey2015}. The ICRF forms the underlying basis for
positional astronomy as well as for deep
space navigation and has applications in determining the Earth's precise
orientation in space, in monitoring variations in the length of day,
in defining a terrestrial reference frame, in measuring tectonic plate motions, 
and providing phase referencing calibrators for VLBI imaging
and relative astrometry measurements of parallaxes and proper motions.
While ICRF2 had a single catalog of 3414 compact
extragalactic source positions at the standard S/X ($\sim$2.3/8.4 GHz) 
geodetic/astrometric frequencies, ICRF3 contains reference frame catalogs
at three radio frequencies. The primary ICRF3 catalog contains 4536 sources 
at S/X frequencies, with secondary catalogs of 824 sources 
at K band ($\sim$24 GHz) and 678 sources at X/Ka band ($\sim$8.4/32 GHz). However, 
dedicated VLBI astrometric campaigns at all three ICRF3
frequency bands have continued into the present time, adding
many additional sources and improving the overall precision.

 At radio frequencies, many quasars exhibit extended emission, often composed of a
jet emanating from a compact component, the radio core. The radio core is 
believed to be associated with a giant black hole at the center of a quasar or galaxy. 
Both the core and jets can be variable and 
VLBI measurements of source positions can vary with 
time, frequency and baseline projection, which can
introduce significant variations in the VLBI
measurements, thereby degrading the accuracy of the estimated source
positions \citep{charlot1990} and limiting the stability of the 
celestial reference frame (CRF).
The higher the frequency and resolution, the more separated are
the extended structure from the radio core.
At K band, quasars generally exhibit more compact source structure 
and reduced core-shifts \citep{charlot2010} than at S/X band. Thus, astrometric VLBI 
observations at K band may permit the construction of a 
more stable and accurate celestial reference frame than at S/X band, which would also be 
advantageous in tying optical reference frames such as 
Gaia \citep{mignard2018, {lindegrin2021}} to the VLBI reference frame.

\begin{figure}[htb!]
         \centering
         \includegraphics[width=0.45\textwidth]{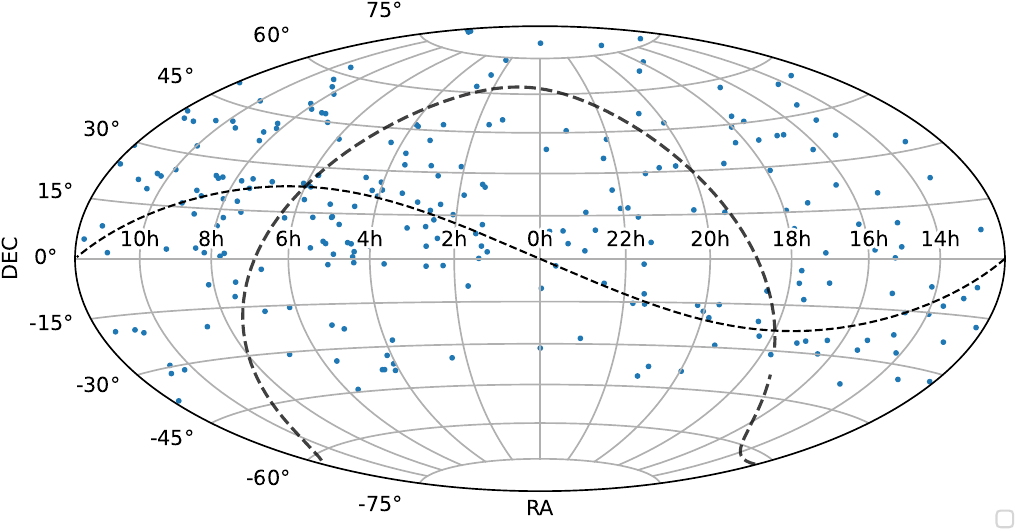}
         \caption{Distribution of the 268 K band sources from the ten
          24-hour VLBA sessions of \citet{lanyi2010}. The dotted 
          lines show the ecliptic and the galactic plane.} 
         \label{fig:sources1}
\end{figure}

An initial K band reference frame was constructed some 16 years ago 
from 10 Very Long Baseline Array (VLBA) sessions taken between 2002 and 2007 
\citep{lanyi2010}. This initial catalog was limited to only 268 sources
and had weak coverage in the mid-south and no coverage in the far south 
($\delta < -45^\circ$). It also had several localized regions with no 
sources, especially near the galactic plane.
Figure~\ref{fig:sources1} shows the sky 
distribution of these 268 sources.
For K band to be developed into a serious celestial reference frame, 
additional observations of sources in both hemispheres  
were needed to densify the frame and obtain full sky coverage.
Thus, in 2014, we began a new series of dedicated astrometric
VLBI sessions at K band in the southern hemisphere using the 
HARTRAO 26-meter antenna at the Hartebeesthoek site of the South 
African Radio Astronomy Observatory (SARAO) and the HOBART26 
26-meter antenna in Tasmania, Australia.
And we resumed northern hemisphere astrometric and imaging observations 
with the VLBA in 2015.
These sessions, through early 2018, were used to construct the
ICRF3-K catalog \citep{charlot2020}. These K band sessions have continued 
into the present time, and more recently, we have added K band observations 
from other networks in both the northern and southern hemispheres. More 
details on the K band project can be found at the K band astrogeo web 
site\footnote{\url{https://sites.google.com/sarao.ac.za/k-bandastrogeovlbi}}.

ICRF3 was approved as Resolution B2, 
2018\footnote{\url{https://drive.google.com/file/d/
\\1i4HdhC\_3v4o0roR3RKUTb8fnvW2Rl0rh/view}}
at the 2018 IAU General Assembly. Paragraph 9 of that resolution states
`that the organizations responsible for astrometric and geodetic VLBI 
observing programs (e.g. IVS\footnote{International VLBI Service for Geodesy 
and Astrometry \citep{nothnagel2017}}) take appropriate measures to continue and 
develop such programs, at multiple radio frequencies and with a specific 
effort on the southern hemisphere, to both maintain and improve ICRF3.'  
The K band CRF was basically in its infancy at the time of ICRF3 and
our K band dataset now contains over seven times as much
data as was used for ICRF3-K, with greatly improved precision and
an additional 493 sources, of which 23 are south of -40$^\circ$ 
declination. Therefore we are presenting here an updated K band catalog,
designated as CRF-K-2025. CRF-K-2025 is a complete revision of all K band 
sources that were in ICRF3-K as well as the additional 493 sources, 
and thus is not simply an extension of \mbox{ICRF3-K}.

\section{Observations} \label{sec:observations}

Building a modern celestial reference frame with VLBI is a major
undertaking typically requiring many years of observations. We 
inherited 15 K band VLBA sessions from earlier 
programs. This included the 10 24-hr sessions from \citet{lanyi2010} 
(projects BR079, BL115, and BL122 from May 2002 - March 2007) plus two
later sessions from the same campaign (BL151 in 
2008), all run at a data rate of 128 Mega-bits-per-second (Mbps). 
We also added the original processing at the Goddard Space Flight Center
of the three galactic plane survey sessions [described by \citet{petrov2011} 
(VLBA project BP125 in 2006)] run at 256 Mbps. Some 25 additional 
VLBA sessions from two new programs, projects BJ083 (4 sessions from
December 2015 - June 2016) and UD001 (21 sessions from January 
2017 - May 2018) and 16 sessions from our southern hemisphere 
campaign using the HARTRAO-HOBART26 baseline (projects ks14, ks16, 
ks17, and ks18 from May 2014 - April 2018) were added for the 
ICRF3-K solution \citep{charlot2020}. One of these southern sessions 
also included the Tidbinbilla 34 meter antenna 
in Australia and another included the Tianma 65 meter antenna in China.
Since the cutoff date for the ICRF3-K solution, we have added
an additional 167 VLBA
databases at 93 epochs, with the latest data used here taken in December 2025.
And we have added 23 additional HARTRAO-HOBART26 databases,
with the latest data taken in February 2021. We also obtained one
session (VT29A) in July 2020 using 5 sites in Australia [the Australian 
Telescope Compact Array (ATCA), CEDUNA, HOBART26, MOPRA 
and PARKES], plus HARTRAO and SEJONG (South Korea). We have also added 8 databases
at 4 epochs from May 2022 - January 2023 using the HARTRAO-YEBES40M (Spain)
baseline. And lastly we have added 34 databases from March 2023 - May  
2025 using the Korean VLBI Network (KVN). The KVN sessions used between 2 and 4 of
the 21 meter KVN antennas (KVNYS, KVNUS, KVNTN, KVNPC) plus other stations
(hereafter referred to as KVN$^{+}$ sessions). The other stations have been
HARTRAO in all 34 sessions, MOPRA in 16 sessions, YEBES40M in 15
sessions, and SEJONG in 8 sessions. Beginning in 2015, our VLBA 
sessions recorded the right circular polarization (RCP) signal at 2 
Gbps, giving a nearly four-fold increase in sensitivity over the 
\citet{lanyi2010} observations. And since November 2019, we have made 
and analyzed 74 VLBA sessions in which both the RCP and the LCP  
signals were recorded at a bit rate of
4 Gbps. Each of these was processed as separate RCP and LCP databases, 
giving the same sensitivity as the earlier VLBA 2 Gbps RCP sessions. 
The 4 HARTRAO-YEBES40M sessions also recorded both RCP and LCP signals at 4 Gbps 
and were processed as separate RCP and LCP databases. 
The KVN$^{+}$ sessions recorded the RCP at 2 Gbps at all stations
except MOPRA which recorded at 1 Gbps.
Data rates for the southern hemisphere sessions were initially 1~Gbps 
for the first two sessions and then were increased to 2~Gbps in 2016.
For this work we used a total of 289 astrometric 
databases taken at 211 epochs. The VLBA and KVN sessions were designed
for both high precision astrometric positioning of the sources as well 
as for imaging. This report concentrates on the astrometry.
Imaging of the VLBA sessions from 2015 to 2018 is reported in \citet{dewitt2023}.

Sources for these sessions were selected based on their flux strength
(typically \(> 200\) mJy) and compactness on milli-arc-second (mas)
scales. The source list 
was formed by starting with sources from the successful
campaigns of \citet{lanyi2010} and \citet{petrov2011} and we added
sources in the far south that had been successfully observed at X/Ka band.
We also added sources that were known to be compact on mas
scales through images at S/X band, or that were known to be
flat spectrum sources. The source list also included ICRF2 and ICRF3
defining sources to provide strong ties to the ICRF. More recently, 
we added some 286 sources from the second Korean VLBI Network Calibrator 
Survey (KVNCS2) \citep{lee2023}.
For VLBA schedules, we adopted a core 
list of $\sim$40 sources and spread out the remaining $\sim$1200 sources 
over a rotation of 6 sessions. 
VLBA schedules were written using the dynamic mode of the NRAO 
SCHED\footnote{\url{http://www.aoc.nrao.edu/software/sched/index.html}}
program, allowing them to be run at any time.
The goal was to create schedules that enabled
both accurate astrometry and sub-mas resolution imaging.
These goals require obtaining several scans on each source spread
through the 24 hour session so as to sample a wide range of hour 
angles and elevations needed for good astrometry and to
obtain good UV coverage for imaging.
We typically obtained 2--3 scans per source in each session.
In January 2017, we began running approximately monthly K band 
VLBA sessions under the USNO's VLBA time allocation (project codes
UD001, UD009, UD015, UD017, UD019 and UD020) in support of USNO's ongoing 
research into the celestial reference frame and geodesy.
The HARTRAO-HOBART26 schedules were also written using SCHED.
In these schedules, typically $\sim$125 sources were observed 3-5
times each, with most of the sources being south of 
$-30^\circ$ declination. However, many southern sources were
not detected due to the lower sensitivity on this long single baseline.
The HARTRAO-YEBES40M sessions and the KVN$^{+}$ sessions were also scheduled
using SCHED, and these allowed longer north-south baselines than the VLBA
sessions, as well as spanning the northern and southern hemispheres,
which can potentially improve the declination precision of sources
previously observed only with the VLBA.
The vast majority of the observations (98\%) have come from 
the VLBA sessions, which have a southern cutoff at around -46.5$^\circ$ 
declination. However, the small percentage of sessions from the 
southern hemisphere has allowed us to include 109 sources south
of $-46.5^\circ$, thus providing full sky coverage for CRF-K-2025.
Of the 1317 sources presented here, 1208 have been 
observed by the VLBA and 377 have been observed from the  
HARTRAO-HOBART26 baseline and the 7-station southern hemisphere 
session, with 268 
sources being observed successfully on both networks in the declination range from
+39.8$^\circ$ to $-46.8^\circ$.

This work involves a significant increase in data over ICRF3-K. The number 
of VLBA sessions used has increased from 40 databases at 40 epochs to 207
databases at 133 epochs, the number of VLBA observations has increased
nearly seven-fold ($\sim$479,000 vs. $\sim$3.45 million) and the number of sources 
detected by the VLBA has increased from 729 to 1208.
For the southern hemisphere effort, the number of sessions has 
increased from 16 to 40, the number of observations has nearly tripled
(from 3671 to 10,464) and the number of sources detected 
with 7 or more observations 
has increased from 279 to 344. More recently, the four HARTRAO-YEBES40M
sessions have obtained 2229 observations and observed 355 of the sources.
And the 34 recent KVN$^{+}$ sessions have added 74,271 observations and
observed 939 of the sources to as far south as -81$^\circ$ declination.
In total, the number of delay observations has increased seven-fold, 
from 482,616 of \mbox{ICRF3-K} to 3,534,340.

\section{Analysis} \label{sec:analysis}

Correlation of all K band VLBA sessions since 2015 used the DiFX correlator
\citep{deller2011} at the National Radio Astronomy Observatory (NRAO)
in Socorro, New Mexico. The earlier VLBA sessions used the now retired VLBA hardware correlator.
Southern hemisphere HARTRAO-HOBART26 sessions were correlated using the 
DiFX correlators at either the Bonn correlator (Germany) or at the 
Hobart correlator (Tasmania, Australia). HARTRAO-YEBES40M sessions were correlated 
on the DiFX correlator at Yebes Observatory (Spain) and the KVN$^{+}$ sessions were
correlated on the DiFX correlator of the KVN.  
The VLBA and KVN$^{+}$ DiFX outputs were converted to both Haystack 
Mark4 format files for astrometry and FITS-IDI format files for imaging,
while the HARTRAO-HOBART26 and HARTRAO-YEBES40M outputs were converted only 
to Mark4 format since imaging normally requires at least 4 stations.
Fringe fitting to obtain the geodetic observables (group delays and phase
delay rates) was performed at either the Bonn correlator, the Hobart correlator,
the Goddard Space Flight Center or the U. S. Naval Observatory using the 
Haystack Observatory 
\textit{HOPS}\footnote{\url{https://www.haystack.mit.edu/haystack-observatory-postprocessing-system-hops/}} 
package and the outputs were 
converted into geodetic-style VLBI databases. Astrometric analysis of each 
session was made at either the U.S. Naval Observatory (USNO) or 
at the Goddard Space Flight Center (GSFC) using the 
\textit{Calc/Solve/nuSolve} analysis package \citep{ma1986, charlot2020, bolotin2014}.

For CRF-K-2025, a least squares solution was made at USNO using 
group delays from all 289 astrometric databases. In the least squares 
solution, we solved for global source coordinates, antenna site positions, 
and most antenna site velocities. Earth Orientation Parameters (EOP's) 
were adjusted from the USNO Bulletin A EOP series in the VLBA and KVN$^{+}$ 
sessions, but not in the single baseline sessions, where there is 
insufficient information to solve accurately for EOP's.
Geophysical models in accordance with the IERS Conventions (2010) 
\citep{petit2010} were applied. 
We applied atmospheric delays from the Vienna Mapping Function 1
model (VMF1) \citep{boehm2004} for sessions before 2008 and  
from the Vienna Mapping Function 3 model (VMF3) \citep{vmf3} for 
later sessions and residual atmospheric delays were solved for.

 Alignment with ICRF3 was made via a no-net-rotation (NNR) 
constraint using 280 of the 303 ICRF3 defining sources. 
The other 23 ICRF3 defining sources (all in the southern part of the
sky) were either not observed
at K band or were not observed enough to be used in the NNR constraint.
This represents an improvement over ICRF3-K, where only 186 
ICRF3 defining sources were used.
This alignment prevents the likelihood of any significant systematic
offsets between the various ICRF catalogs. 
Some 3.53 million
individual baseline observations were used for the CRF-K-2025 solution. 
This is in contrast to the $\sim$13 million S/X observations of ICRF3-SX
\citep{charlot2020}, but over seven times as many observations as were
used for ICRF3-K.

\begin{figure}[htb!]
         \centering
            \includegraphics[width=.45\textwidth]
           {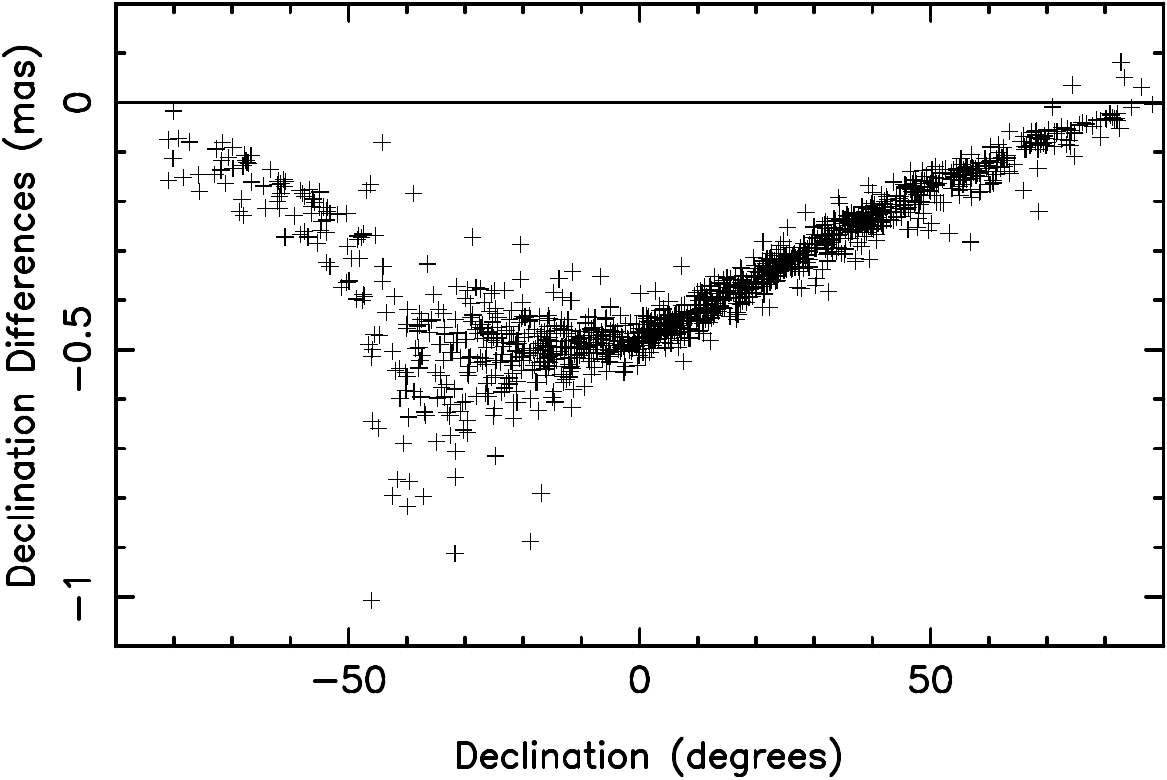}
         \caption{The effect of including GNSS ionosphere calibrations on 
          declination at K band.
          Declination with GNSS ionosphere calibrations minus declination
          without GNSS ionosphere calibrations is plotted versus declination.
          The effect on right ascension (not shown) is negligible.}
         \label{fig:gps}
\end{figure}

Delays due to the ionosphere were computed as described in \citet{lanyi2010} and 
\citet{charlot2020} using total electron content (TEC) maps produced by the 
Jet Propulsion Laboratory \citep{martire2024} 
using data from global navigation satellite systems (GNSS). These maps are given at 
2-hour intervals and have a resolution of 2.5$^\circ$ by 5$^\circ$ in 
latitude and longitude, respectively. This temporal and spatial resolution
cannot provide the accuracy of simultaneous dual-frequency observations, 
but does allow averaging out of most of the systematic effects of the ionosphere,
provided a sufficient number of observations are obtained on each source.
Accurate ionosphere calibrations are necessary to avoid declination zonal errors.
In Figure~\ref{fig:gps} we show the effect of applying GNSS
ionosphere calibrations, where the difference in source positions 
determined with and without the GNSS ionosphere delays applied is plotted versus 
declination. Figure~\ref{fig:gps} shows a maximum effect of $\sim$500 $\mu$as
just south of the celestial equator. The greater scatter starting at
around $-10^\circ$ declination we believe is due to a combination of the VLBA observing at 
lower elevations and with fewer antennas, the noisier data from the 
southern network, and the fact that there are fewer GNSS stations in the 
southern hemisphere to contribute to the ionosphere maps. 
On average, the ionospheric calibrations have negligible systematic affects on  
source right ascensions. 

Some challenges were encountered in combining the data from the different
networks. The vast majority of the data comes from VLBA-only sessions,
with no direct connections at K band to any of the sites in the other networks.
However, we do have ties between the VLBA stations and HARTRAO, HOBART26 
and YEBES40M at S/X band from the IVS RDV/RV sessions. And 
HARTRAO has participated in all of the non-VLBA K band sessions so far.
This allows us to use HARTRAO as the reference site in those sessions by 
fixing it to its S/X position and velocity. Thus we are able 
to stitch together the various disjoint networks fairly successfully.
Overall source precision is dominated by the VLBA geometry over $\sim$2/3
of the sky, where the north-south 
extent of the VLBA network is considerably less than its east-west extent, 
resulting in lower source precisions in declination than in right ascension.
Among the non-VLBA sessions, there are some baselines crossing the equator 
with longer north-south extent (most notably the HARTRAO-YEBES40M baseline)
that have been used in several sessions and these have helped to improve 
the declination precision by a small amount. In 
stitching together source positions from the various 
networks, some rotations and/or declination zonal errors are possible. 
However the overlap of 268 sources between +39$^\circ$ and 
$-46^\circ$ declination between the VLBA and the southern hemisphere sessions 
should help to limit any such distortions.
However, we do see a small rotation of $\sim$8 $\mu$as about the Y-axis 
(which passes through the equator at 6 hours right ascension) and smaller 
rotations about the X and Z axes of our initial solution with respect to 
ICRF3-SX. Therefore, to align our CRF-K-2025 solution
with ICRF3, we imposed a very small solid-body rotation, 
similar to how ICRF2 was aligned with the first ICRF \citep{fey2009}.

In the least-squares solution method, the formal uncertainties
estimated for source positions will decrease as the square root
of the number of observations 
once the full range of geometry has been observed.
The formal uncertainties can thus become
unrealistically small for heavily observed sources as the number of
observations becomes very large. Additionally, not all of the observations from
a VLBI network are independent. For example, a scan using the 10 station
VLBA network will yield 45 observations, but only 10 are fully independent.
The least squares method also does not fully take into account source
`jitter' \citep{cigan2024}, which is caused by both instrumental limitations,
source structure variations, limitations in estimating the atmospheric
and ionospheric delays, variations in the observing networks, and
other factors. 
\citet{ryan1993} first investigated the validity of the formal
VLBI uncertainties coming from the least-squares analyses and concluded
that a scaling factor of 1.5 should be applied to VLBI formal
uncertainties to bring them closer to actual uncertainties.
This 1.5 scaling factor was applied in all three ICRF VLBI realizations
\citep{ma1998, fey2009, fey2015, charlot2020}.
Also in the three ICRF VLBI realizations, noise floor analyses were
made to estimate the minimum realistic noise levels. For
ICRF3-K, the dataset was divided into two parts and independent
solutions were made using half the databases in each solution. The noise
floors were estimated in 10 degree declination bands as the WRMS
of the source differences divided by the square root of 2. At that
time, the K band dataset was much smaller than now, and the analysis
was quite limited in the southern part of the sky. Although differences
were seen between the north and the south, single values of 30 and 50
$\mu$as were adopted for the noise floors in right ascension
and declination, respectively.
The asymmetry in the noise floors was due to the fact that most of the
data came from the VLBA, where the
east-west extent is considerably greater than the north-south extent.
We now have much more K band data
and have repeated this type of analysis covering the full declination
range, although the data is still sparse in the far south and most of
the data still comes from VLBA sessions. The average
values found are approximately 47 and 78 $\mu$as
in RA and declination.
These are greater than what was seen for ICRF3-K,
perhaps due to the longer time span allowing more 'jitter' to occur
and perhaps to greater solar activity affecting the ionosphere over the last few years.
The comparisons also show a clear declination dependence, which we now decide
to include. The estimated noise floors range from 37 to 130 $\mu$as
in RA and from 60 to 140 $\mu$as in declination. Table 1 gives the RA and declination
noise floors at 20 declinations, and we interpolate linearly in declination to compute
the noise floor for each source individually. Accordingly then, we have 
re-scaled the formal errors by a factor of 1.5 and then added the interpolated 
noise floor values in quadrature.

Some sources have shown significant changes over time in their solved-for positions 
at S/X and/or K bands. Brightness variations in their structure along a jet or the core 
can usually account for such apparent position changes. 3C48 (0134+329, J0137+3309)
is a prime example \citep{titov2022} (see 'Notes on Several
Sources' section), showing a jump of $\sim$56 mas between 2017 and 2018 at X/S band.
We compared source position differences between CRF-K-2025 and ICRF3-K. 
For a more realistic comparison, the ICRF3-K inflated source
uncertainties were recomputed using our current declination
dependent noise floor estimates. Of the 824 common sources, 16
have right ascension and/or declination differences
exceeding three times the combined inflated uncertainties,
with 5 (3C119, 0057+678, 0723-008, 0743-006 and 1600+335) exceeding 4 
times the combined uncertainties. We have looked at these 5 sources in detail.
For 3C119 (0429+415, J0432+4138), we get an offset of 
$\sim$6.8 mas to the SW from the ICRF3-K position. VLBA images
\citep{mantovani2010} at 4.8 and 8.4 MHz show considerable structure with  
three components extending over $\sim$40 mas in a NE-to-SW line. The indication is 
that this offset is due to brightness variations along this structure.
1600+335 (J1602+3326) shows an offset of $\sim$0.77 mas from ICRF3-K to 
the NNE and K band VLBA images from \citet{dewitt2023} 
show a jet to the NNE with 2 components and an extent of $\sim$3 mas.
0743-006 (J0745-0044) shows an offset from ICRF3-K of $\sim$0.57 mas to the SW.
VLBA images at 15 and 43 GHz \citep{Cheng2020, Cheng2023} and K band \citep{dewitt2023} 
show it to be elongated with 2 main components $\sim$2 mas apart in the 
NE-to-SW direction.
0723-008 (J0725-0054) shows an offset of $\sim$0.67 mas to the NNW from ICRF3-K.
VLBA images at 43 GHz \citep{Cheng2020} and K band \citep{dewitt2023} 
both show two components in a SSE-to-NNW line extending
over $\sim$2--3 mas.
0057+678 (J0100+6808) shows an offset of $\sim$1.54 mas from ICRF3-K to the NNW.
Several K band VLBA images by \citet{dewitt2023} show it to be 
elongated over $\sim$2 mas in the SSE-to-NNW direction.
These five cases are all consistent with apparent motion towards 
or away from a jet, indicating structure-brightness variations
which effects the average observed source position 
and is something which the least squares
solution and the noise floor estimates cannot easily capture
and which do not represent the overall intrinsic precision of the CRF-K-2025
catalog. Each of these sources appear to be stable at the currently estimated
coordinates so we include them in CRF-K-2025. 
For all common sources, the total position differences from ICRF3-K vary from 
0.001 mas to 6.78 mas, with a median value of 0.21 mas. 
A similar comparison of CRF-K-2025 with ICRF3-SX using 1277 matched sources 
gives position differences ranging from .005 mas to 34.81 mas,
with a median offset of 0.35 mas. This compares to ICRF3-K vs ICRF3-SX
where the differences range from .004 mas to 35.00 mas, with a median value of .36 mas.

\begin{table}
 \caption{K band noise floor values estimated in right ascension and declination
 as a function of declination.                                           }
 \label{tab:noise    }
\begin{tabular}{lll}
   Dec. &     RA ($\mu$as)   &  Dec ($\mu$as)  \\
   \hline
    --90.       &    130   &  140  \\
    --85.       &    130   &  140  \\
    --75.       &     95   &  135  \\
    --65.       &     90   &  135  \\
    --55.       &     80   &  140  \\
    --45.       &     80   &  150  \\
    --35.       &     75   &  150  \\
    --25.       &     45   &  115  \\
    --15.       &     37   &  100  \\
    --05.       &     37   &   90  \\
    +05.       &     36   &   70  \\
    +15        &     37   &   70  \\
    +25.       &     45   &   70  \\
    +35.       &     50   &   70  \\
    +45        &     50   &   60  \\
    +55.       &     50   &   60  \\
    +65.       &     50   &   60  \\
    +75        &     50   &   60  \\
    +85        &     55   &   60  \\
    +90.       &     55   &   60  \\
\hline
\end{tabular}
\end{table}

The sun's motion around the galaxy produces a large shift in the
apparent positions of the quasars due to aberration (of up to $\sim$160 arc seconds) 
and the slow change in the sun's direction will produce a slow change in
these aberrated positions. 
Sometimes called `secular' or `galactic' aberration, this
effect will be manifested as a streaming motion in a dipole pattern towards 
the direction of solar acceleration, 
which is presumed to be in or near the direction of the galactic center.
As was done with ICRF3, our CRF-K-2025 solution was made by applying a solar 
acceleration dipole model, as
described in \citet{macmillan2019} and \citet{charlot2020}, using a solar 
acceleration constant of 5.8 $\mu$as/yr in the direction of the galactic center.
To clarify, source positions are 
determined in the J2000 reference frame, but at a solar acceleration epoch 
of 2015.0, as was done for all three catalogs of ICRF3 \citep{charlot2020}.
Following the nomenclature of \citet {titov2010} and \citet{liu2012}, the
secular aberration motion per year for a source in right ascension and declination 
(in $\mu$as/yr) is:
\begin{displaymath}
\Delta\alpha = 5.8 (-d1 \times sin(\alpha) + d2 \times cos(\alpha))/cos(\delta) 
\end{displaymath}
\begin{displaymath}
\Delta\delta = 5.8(-d1\times sin(\delta)cos(\alpha) - d2\times sin(\delta)sin(\alpha) + d3 \times cos(\delta)) 
\end{displaymath}
Where
d1 = -0.05467006, 
d2 = -0.87284157, and   
d3 = -0.48493173 are the components of the acceleration unit vector in the direction of the 
galactic center. For the most accurate positions at other epochs, multiply 
these offsets by the number of years since 2015.0 and add to the CRF-K-2025 coordinates.

We are requiring at least 7 observations of a source
for inclusion in the CRF-K-2025 catalog. Sources 
with fewer observations have large uncertainties and 
their solved-for positions can change significantly 
when additional observations are obtained, this probably being more likely 
for single band observations where GNSS ionosphere corrections are applied.
Thus a handful of sources with fewer than 7 baseline
observations were excluded, all observed only from the southern
hemisphere HARTRAO-HOBART26 baseline. This is in 
contrast to ICRF3-K, where 5 sources with less than 7 observations 
were included. However, all 824 sources from ICRF3-K now have sufficient
observations to be included in CRF-K-2025.

During the course of our K band campaign, SGR A*, the supermassive black
hole in our galaxy, was also observed in numerous VLBA sessions.
Since it is not an extragalactic source, SGR A* is not included in 
CRF-K-2025. However, those observations led to the determination of
the position and proper motion of SGR A* in the ICRF3 frame \citep{gordon2023}.

\section{Astrometric Results} \label{sec:results}

\begin{figure}[tbh!]
         \centering
         \includegraphics[width=0.45\textwidth]
         {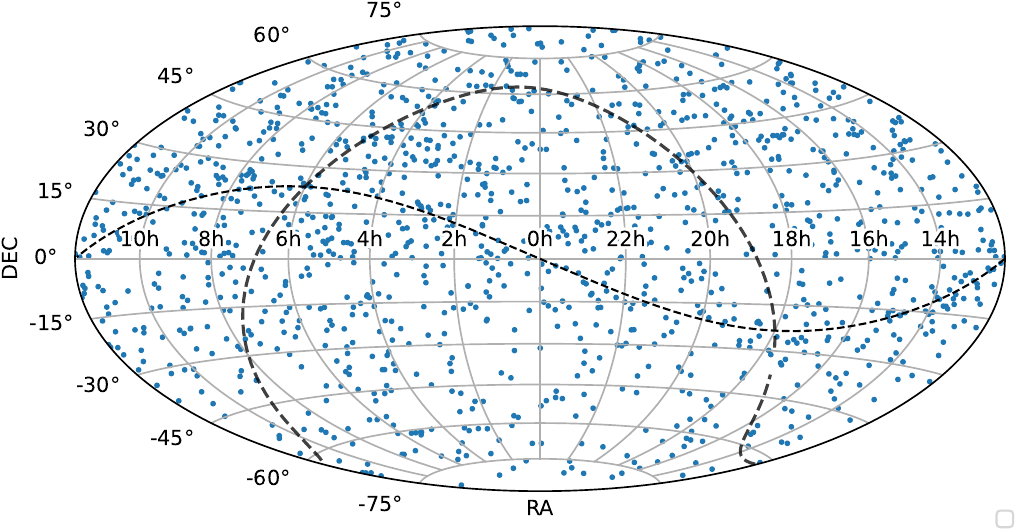}
         \caption{Distribution of the 1317 sources of the \mbox{CRF-K-2025}
	 catalog. The dotted lines show the ecliptic and the galactic plane.}  
         \label{fig:sources2}
\end{figure}

\begin{figure}[bht!]
         \centering
         \includegraphics[width=.45\textwidth] 
           {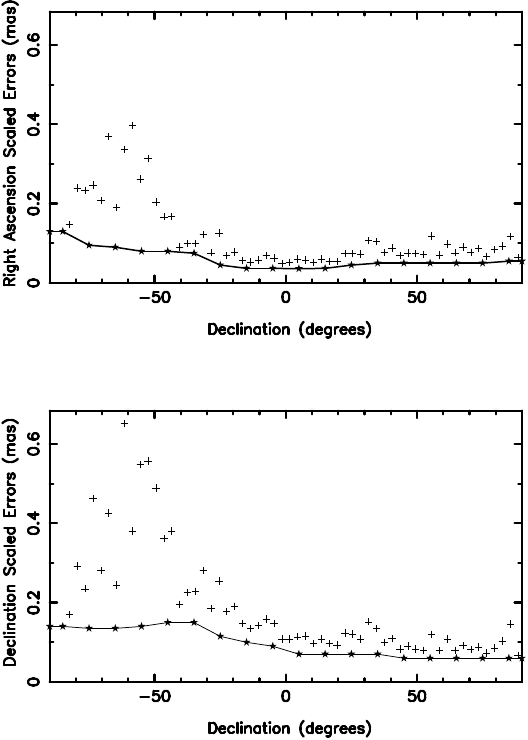}      
         \caption{The distribution of scaled uncertainties averaged in $3^\circ$ declination bins
         for CRF-K-2025 (plus signs) and the estimated noise floor values (connected lines). }
          \label{fig:uncer}
 \end{figure}

We now have 1317 extragalactic sources in the CRF-K-2025 catalog detected in
at least seven baseline observations and covering the full sky.
Figure \ref{fig:sources2} shows the distribution of these 1317 sources.   
The distribution of the scaled position uncertainties is shown in 
Figure~\ref{fig:uncer},
averaged in 3$^\circ$ declination bins. The precision in right ascension 
begins slowly degrading at around \mbox{$-20^\circ$} declination where 
we begin losing the longest East-West VLBA baselines, and then rapidly 
south of around $-45^\circ$, which is the approximate
limit of the VLBA and where we get observations only from the 
southern hemisphere baselines. In declination, the precision begins degrading 
south of around $-5^\circ$, where we begin to lose the two northern-most VLBA 
antennas (HN-VLBA and BR-VLBA), and then more rapidly as we get only
the southern hemisphere observations. 
Overall median scaled formal uncertainties are 60 $\mu$as in 
right ascension and 104 $\mu$as in declination.

We have made comparisons of the precision between our new CRF-K-2025 catalog 
versus ICRF3-K, ICRF3-SX, the latest USNO S/X solution and a 
recomputed version of the \citet{lanyi2010} catalog.
The median and average inflated uncertainties for the sources in common 
in these comparisons are listed in Table 2.
These results indicate that, for the 824 ICRF3-K sources,
CRF-K-2025 is $\sim$25\% more precise than 
ICRF3-K, even using the larger noise floor values. And it is 
$\sim$15\% more precise in right ascension but slightly noisier 
in declination than ICRF3-SX for 1277 sources in common with 10 or more observations. 
However, comparison with the latest S/X 
catalog\footnote{see \url{http://crf.usno.navy.mil/quarterly-vlbi-solution}} 
at USNO being developed for ICRF4 
(and rescaled as for ICRF3-SX), shows CRF-K-2025 to be
noisier by roughly 10 and 40 percent 
in right ascension and declination, respectively, for 1302 common sources.
We also ran a solution on the original 10 K band sessions used for the catalog of
\citet{lanyi2010}, applying the current geophysical models, software, and error scaling
for comparison to our current K band catalog. This comparison
shows very significant improvement in K band precision
since those first epoch K band observations.
%

\begin{table}
\begin{center}
 \label{tab:precision}
 \caption{Median/Average inflated uncertainties in various comparisons. 
  Column~1 lists the band and catalog. Column~2 lists the number of sources or 
  the number of common sources between catalogs.
  `Latest-X/S' is the current S/X catalog at USNO.
  The `K-2010*' catalog is a re-analysis of
  the 10 sessions of \citet{lanyi2010} using current geophysical models
  and with errors re-scaled as in CRF-K-2025. }
\begin{tabular}{lccc}
                               
 Catalog &~\# Sources~    & RA $\sigma$ [$\mu$as] ~ &  Dec $\sigma$ [$\mu$as]~  \\                       
 \hline
  CRF-K-2025 (all)       & 1317  & 60/92 & 104/158  \\   
 \hline
  CRF-K-2025             & 824    & 56/84  & 95/145  \\   
  ICRF3-K                & 824    & 73/139 & 135/258 \\   
 \hline
  CRF-K-2025             & 1277    & 60/86  & 103/149   \\  
  ICRF3-SX               & 1277    & 71/104  & 100/161  \\  
\hline
  CRF-K-2025             & 1302    & 60/88  & 104/152  \\   
  Latest-S/X             & 1302    & 54/73  & 74/105   \\   
 \hline
  CRF-K-2025             & 268    & 50/52   & 82/94       \\   
  K-2010*                & 268    & 123/541 & 242/1065    \\   
 \hline

\end{tabular}
\end{center}
\end{table}

CRF-K-2025 provides various improvements over ICRF3-K, but still has
some deficiencies. It has an additional
493 sources ($\sim$60\% more) with 23 of those being south
of $-40^\circ$ declination, increasing the number of `southern' sources by
$\sim$20\% (134 vs 111).
And it was generated from $\sim$3.53 million individual baseline observations, 
over 7 times as much data as was used for ICRF3-K. 
Data from the
southern hemisphere networks has increased by a factor of 2.3
since ICRF3-K, but is still only
a very small fraction of the total data, resulting in an asymmetric distribution
of uncertainties, as can be seen in Figure~\ref{fig:uncer}.
The distribution of sources though, is not so asymmetric
with $\sim$60\% of the sources north of the celestial equator
and $\sim$40\% south of it.
The lack of direct connections between the VLBA and the smaller 
northern and southern networks is of some concern, as it could
possibly introduce distortions and/or rotations. It was desired to align
CRF-K-2025 with the ICRF3 defining sources. However, 23 of the ICRF3 
defining sources
were either not observed at K band or not observed enough to be used
for a proper alignment. Thus we made the alignment with 1211 
ICRF3-SX sources by applying a small solid-body rotation 
(of -0.9, 7.9, -3.5 $\mu$as in X,Y,Z) to the initial solution
to produce CRF-K-2025. 

To further examine the possibility of
various distortions, we have made comparisons of CRF-K-2025
with other catalogs using the method of vector spherical 
harmonic (VSH) decomposition developed by \citet{mignard2012}, as was done for 
ICRF3 \citep{charlot2020}. To avoid biasing the results from
large outliers, we excluded sources in which the position differences
were greater than 5 times the combined uncertainties, or the position
differences were greater than 5 mas, or there were less than 10 observations
of the source in either catalog. Table 3 lists the 16 VSH components (in $\mu$as) 
of the comparisons of CRF-K-2025
with common sources in the ICRF3 defining list, ICRF3-SX, ICRF3-K, 
and Gaia-DR3 \citep{gaia2016,gaia2023}. 
With respect to ICRF3-SX, there are significant differences in 2 of
the dipole terms and in several of the quadrupole terms, however
none are excessively large. The comparison with ICRF3-K shows 
small rotations and also some significant dipole
terms, but none being excessive. In the comparison to Gaia-DR3, 
none of the rotations or dipole terms are significant, but some of
the quadrupole terms are significant, though none are excessive.
Overall, we believe these
various comparisons are reasonable and any differences between catalogs are not
excessively above the noise levels. 
It should also be pointed out that both ICRF3-SX and ICRF3-K also
show significant dipole and quadrupole differences compared to Gaia-DR3
and that CRF-K-2025 shows considerably better agreement with Gaia-DR3 than does 
ICRF3-K.

The current observational deficiencies are not likely to improve in the short term.
However, a set of 10 planned European VLBI Network (EVN) global K band 
sessions with up to 28 stations in both the northern and southern hemispheres
over the next 2--3 years should directly connect the VLBA with the other stations.
Also in the longer term, possible use of new antennas
in Thailand and Argentina could have a significant impact. 
Use of the new 40 meter antenna of the Thai National Radio Astronomy
Observatory\footnote{\url{https://www.narit.or.th/en/observatory/thai-national-radio-telescope}}
 \citep{sugiyama2024},
at $+19^\circ$ latitude, could help to densify and improve 
K band precision down to around 
$-60^\circ$ declination
when used with southern hemisphere antennas and/or in future global sessions.
And use of the still under construction 40 meter China-Argentina Radio Telescope 
(CART) \citep{li2025}, 
at $-32^\circ$
latitude, could potentially densify the entire southern hemisphere
with precisions on par with the northern hemisphere.

\begin{table} [t]
\centering
 \caption{Parameters of the transformations between \mbox{CRF-K-2025}
 and the ICRF3 defining sources, the ICRF3-SX catalog, the ICRF3-K
 catalog, and the Gaia-DR3 catalog, in $\mu$as. The $R_1$, $R_2$ and $R_3$ terms are the 
 solid body rotations in the X,Y,Z directions. The $D_1$, $D_2$
 and $D_3$ terms are the glide (dipole) terms. And the $E_{2i}$ (electrical/poloidal)
 and $M_{2i}$ (magnetic/toroidal) are quadrupole terms representing degree-2 deformations. } 
 \scriptsize{\tiny}
\begin{tabular}{|l|c|c|c|c|}

\hline
              &    Defining      & ICRF3-SX        & ICRF3-K        &   Gaia-DR3      \\
\hline
\# Sources    &       284        &   1211          &     803        &       943       \\
\textbf{Rotation}  &             &                 &                &                 \\
$R_1$           &   7.8$\pm$9.0  &    0.0$\pm$5.7  &  -6.1$\pm$8.1  &   11.0$\pm$7.7  \\
$R_2$           &  -4.6$\pm$7.9  &    0.0$\pm$4.7  &  21.3$\pm$6.6  &  -14.6$\pm$7.0  \\
$R_3$           &  -0.8$\pm$4.8  &    0.0$\pm$2.9  &   3.2$\pm$3.7  &    1.6$\pm$5.3  \\
\textbf{Glide}  &                &                 &                &                 \\
$D_1$           & -29.4$\pm$9.7  &  -24.4$\pm$6.0  & -31.6$\pm$8.1  &  -19.5$\pm$9.3  \\
$D_2$           & -21.4$\pm$7.8  &  -16.1$\pm$5.0  &  24.3$\pm$7.0  &  -14.2$\pm$7.0  \\
$D_3$           &   7.8$\pm$8.8  &   10.5$\pm$5.4  &  25.7$\pm$8.0  &    5.2$\pm$7.1  \\
\textbf{Quadrupole} &            &                 &                &                 \\
$E_{20}$        &  4.8$\pm$10.3  &    5.6$\pm$6.2  &  -9.3$\pm$9.1  &  -40.5$\pm$8.4  \\
$M_{20}$        &  31.1$\pm$6.5  &   33.6$\pm$3.9  &   8.0$\pm$5.1  &   19.8$\pm$6.6  \\
$E_{{21}}^{Re}$ &   9.9$\pm$9.7  &    3.8$\pm$5.8  & -11.2$\pm$8.0  &   -1.1$\pm$8.5  \\
$E_{{21}}^{Im}$ &  30.1$\pm$9.6  &    9.8$\pm$5.9  & -22.3$\pm$8.2  &  -13.3$\pm$8.6  \\
$M_{{21}}^{Re}$ &  -6.8$\pm$9.3  &   -4.1$\pm$5.7  &  12.5$\pm$7.9  &  -46.3$\pm$8.5  \\
$M_{{21}}^{Im}$ &  7.8$\pm$11.4  &   10.8$\pm$7.0  &  -7.6$\pm$9.4  &  -18.2$\pm$11.0 \\
$E_{{22}}^{Re}$ &  -5.4$\pm$3.3  &  -4.6$\pm$1.9   & -10.5$\pm$2.5  &   -4.2$\pm$3.5  \\
$E_{{22}}^{Im}$ &   1.7$\pm$3.2  &   1.2$\pm$2.0   &  -4.9$\pm$2.5  &    0.6$\pm$3.6  \\
$M_{{22}}^{Re}$ & -12.1$\pm$4.8  & -14.9$\pm$2.8   &  -2.0$\pm$4.0  &   -9.9$\pm$4.2  \\
$M_{{22}}^{Im}$ &  -6.0$\pm$4.8  &  -5.3$\pm$2.8   &  -9.8$\pm$4.0  &    3.9$\pm$4.1  \\
\hline
\end{tabular}

\end{table}

\section{The CRF-K-2025 Catalog} \label{sec:catalog}
A small portion of the CRF-K-2025 catalog is presented in 
Table~\ref{tab:kcatalog}.
The full table is available in machine readable form. 
In addition to the source positions
and scaled errors, we also include for most of the sources 
estimated K band fluxes, the object type, its redshift, Gaia G
magnitude, the number of sessions the source was observed in and the
number of observations of the source used in the solution. Object types, Gaia
G magnitudes and redshifts were extracted from the 
OCARS\footnote{\url{http://www.gaoran.ru/english/as/ac\_vlbi/ocars.txt} and
\url{http://www.gaoran.ru/english/as/ac\_vlbi/ocars\_m.txt}}
catalogs \citep{malkin2018}. 

Fluxes were estimated using the following equation:
\begin{displaymath}
    Flux = \frac{1.75 \times SNR \times \sqrt{SEFD_{1} \times SEFD_{2}}}{1.38 \times \sqrt{\#samples}}
\end{displaymath}
where \textit{Flux} is in Jansky's; \textit{SNR} is the group delay
signal-to-noise ratio computed by the \textit{hops/fourfit}
program; \textit{${SEFD_{i}}$} are
the `system equivalent flux densities' of the two antennas, which are equal to
the system temperatures divided by the antenna gain in degrees per Jansky;
\textit{\#samples} is the total number of digitized samples recorded during
the interval used by \textit{fourfit} for the observation; and the factor
1.38 is for the improvement in sensitivity of 2-bit versus 1-bit sampling.
         For sources observed on the VLBA, we used tables of system temperatures 
for each session and antenna gains from the `vlba\_gains.key' 
file\footnote{\url{https://science.nrao.edu/facilities/vlba/calibration-and-tools/caliblogs}}. 
Two flux values are given for most VLBA sources.
The first is the average flux on projected baselines shorter than 1000 km, 
which represents the approximate total flux. The second is the average on 
the longest baselines, which represents the approximate unresolved flux.
For sources observed only from the southern hemisphere, we were able to
use antab tables (which contain antenna gains and system temperatures) from HARTRAO
for the HARTRAO-HOBART26 sessions. For HOBART26, where antab tables 
were not available, we used an estimated SEFD of 1800 Jansky's. For these 
southern sources we give only an estimated flux for the longest baselines. 
We caution that these flux values
should be considered only rough estimates, as there were no attempts to do
accurate amplitude calibration and they represent the average over many 
different sessions. Further, the \textit{SNR}'s determined by
\textit{fourfit} can be affected by various factors such as phase instabilities
or brief amplitude spikes in the data. Also, flux strengths of these types of
sources can be expected to vary on time scales ranging from a few years to as
little as a few months.


\startlongtable
\begin{deluxetable*}{ l l l l c c c c c c c c c c r r r}
 \centerwidetable
  \label{tab:kcatalog}
  \tablecolumns{17}
\tablecaption{The CRF-K-2025 Celestial Reference Frame } 
 \tablewidth{0pt}
 \tabletypesize{\tiny}
%
 \startdata
J000108.6+191433 & 2358+189 & 2358+189 &     & 00 01 08.6215687 &  19 14 33.801689 &   0.066 &  0.114 &  -0.384 & 60216.7 &  0.083 &  0.064 & AQ & 3.1000 &      &  23 &  1339 \\
J000435.6-473619 & 0002-478 & 0002-478 & Y D & 00 04 35.6554968 & -47 36 19.603937 &   0.189 &  0.408 &   0.592 & 58976.0 &        &  0.130 & AQ & 0.8800 & 19.7 &  29 &    83 \\
J000504.3+542824 & 0002+541 & 0002+541 & Y   & 00 05 04.3633373 &  54 28 24.924357 &   0.062 &  0.071 &   0.001 & 58993.3 &  0.139 &  0.049 & AQ & 0.5040 & 18.6 &  43 &  2975 \\
J000557.1+382015 & 0003+380 & 0003+380 & Y   & 00 05 57.1753856 &  38 20 15.148987 &   0.056 &  0.072 &  -0.255 & 59187.4 &  0.534 &  0.278 & AS & 0.2340 & 18.5 &  55 &  2996 \\
J000613.8-062335 & 0003-066 & 0003-066 & Y   & 00 06 13.8928902 & -06 23 35.335721 &   0.041 &  0.099 &  -0.236 & 59323.3 &  2.110 &  0.434 & AL & 0.3467 & 17.3 &  49 &  3798 \\
J000903.9+062821 & 0006+061 & 0006+061 & Y   & 00 09 03.9318435 &  06 28 21.240050 &   0.057 &  0.114 &  -0.333 & 59479.1 &  0.066 &  0.039 & AL & 1.5626 & 18.0 &  41 &  2756 \\
J000904.1+400146 & 0006+397 & 0006+397 &     & 00 09 04.1735916 &  40 01 46.705523 &   0.128 &  0.158 &  -0.041 & 60332.9 &  0.136 &  0.069 & AQ & 1.8300 & 19.7 &   6 &   279 \\
J001031.0+105829 & 0007+106 & IIIZW2   & Y D & 00 10 31.0059006 &  10 58 29.504291 &   0.043 &  0.081 &  -0.256 & 58471.5 &  0.724 &  0.566 & A1 & 0.0872 & 16.1 &  43 &  3073 \\
J001033.9+172418 & 0007+171 & 0007+171 &     & 00 10 33.9906903 &  17 24 18.761095 &   0.069 &  0.120 &  -0.246 & 60532.1 &  0.243 &  0.122 & AQ & 1.6010 & 16.9 &  12 &   737 \\
J001053.6-215704 & 0008-222 & 0008-222 &     & 00 10 53.6499861 & -21 57 04.219880 &   0.073 &  0.180 &  -0.219 & 60589.9 &  0.170 &  0.076 & AQ & 0.7270 & 20.1 &  13 &  1053 \\
\enddata
 \bigbreak

\tablecomments{Column descriptions: (1) ICRF name. (2) IERS name. (3) IVS name. (4) 'Y' indicates the source was in ICRF3-K, 
'D' indicates it is an ICRF3 defining source. (5) J2000 Right Ascension. (6) J2000 Declination.
(7) RA scaled uncertainty times cosine($\delta$) (mas). (8) Declination scaled uncertainty (mas). 
(9) Correlation between RA and declination. (10) Mean MJD. (11) Total estimated K band flux.
(12) Unresolved K band flux (as seen on the longest baselines).
(13) Type of object.
(AB = blazar, AL = BL Lac object, AQ = quasar, AR = LINER-type AGN,
AS = Seyfert galaxy, A1 = Seyfert 1 galaxy, A2 = Seyfert 2 galaxy,
G = radio galaxy, I = IR source, R = radio source, V = visual source).
(14) Redshift. (15) Gaia G magnitude.
(16) Number of K band VLBI sessions the source has been observed in. 
(17) Total number of K band observations.
Note: types, redshifts and G magnitudes were taken from the OCARS catalogs
\citep{malkin2018}.
}
\tablecomments{This table is published in its entirety in machine-readable
format. A portion is shown here for guidance regarding its form and content.
Currently the CRF-K-2025 catalog can be downloaded from
\url{https://sites.google.com/sarao.ac.za/k-bandastrogeovlbi/data-products/kband-catalog-2025}
}
\end{deluxetable*}

\section{Notes On Several Sources}

3C48B: 3C48 was included in ICRF3-SX using observations from 1990-2017.
Historically it has been used as a flux and polarization calibrator.
However it has a complex radio morphology, see images by \citet{an2010}.
The ICRF3-SX position corresponds to the `B2' component in \citet{an2010}.
It underwent some type of activity between July 2017 and May 2018,
resulting in an apparent displacement of $\sim$56 mas from its ICRF3-SX 
position and corresponding to the `B' component in \citet{an2010}.
The cutoff date for ICRF3-SX
data was March 2018, so the effects of this displacement did not affect 
it's position in ICRF3-SX.
The first K band observations of 3C48 were in January 2019, after this large 
displacement. Thus the CRF-K-2025 position differs significantly
from the ICRF3-SX position, but does agree well with the S/X position 
after May 2018. 
To avoid confusion with the ICRF3-SX position, we are calling this source 3C48B,
although the ICRF and IERS names are not changed.


0218+357 (J0221+355A, aka 0218+35A and 0218+35B) and 1830-211 
(J1833-210A, aka 1830-21A and 1830-21B): 
These two sources are lensed quasars that were excluded from the
ICRF3-SX catalog because a single position could not be solved for at S/X band. 
However, this is not a problem at K band, and these were included in ICRF3-K,
so we also include them in CRF-K-2025. 
However, they should probably not be used for phase referencing. 

\section{Conclusions}

An updated K band CRF catalog is presented here, 
containing an additional 493 sources over ICRF3-K, and with a precision
approximately 25\% better than ICRF3-K. 
At K band, the systematic source structure floor is expected to be 
several times smaller than at S/X band and thus K band has the potential 
to exceed S/X in precision.  
We believe that compared to the current S/X-based IAU standard, the K band 
work represents a very efficient use of resources to achieve a given level of 
astrometric precision while being far less susceptible to astrophysical 
systematics. Work now in progress to realize the full potential of the 
K band CRF includes increasing the temporal resolution of the GNSS ionosphere maps 
to 15 minutes, gathering more southern data, and including future global K band 
sessions with up to 28 stations.
With sources generally being weaker at K than S/X band, the number of
sources in the K band CRF is not expected to approach that of the S/X
catalog. However, the current 1317 K band source catalog
does compare well to S/X in the number of regularly observed sources in
geodetic sessions. We 
are well on our way towards realizing K band's potential to be the basis of
a world class reference frame.

\section*{Acknowledgements}
The VLBA is operated by the National Radio Astronomy Observatory (NRAO), a  
facility of the National Science Foundation and
operated under cooperative agreement by Associated Universities, Inc.
The authors gratefully acknowledge use of the VLBA under the USNO's time 
allocation program since 2017. This work supports USNO's ongoing research 
into the celestial reference frame and geodesy. 
This work made use of the Swinburne University of Technology software
correlator, DiFX, developed as part of the Australian Major National Research
Facilities Programme and operated under license. For a description of
the DiFX correlator, see \citet{deller2011}.
This study also made use of data collected from radio telescopes at
HARTRAO, HOBART26, Yebes Observatory, the Korean VLBI network (KVN), 
the Australian Telescope Compact Array, Mopra, Parkes,
Ceduna, Tidbinbilla, Sejong, and Tianma65. 
The HARTRAO antenna is located at the Hartebeesthoek site and operated 
by the South African Radio Astronomy Observatory (SARAO), a facility of the 
National Research Foundation (NRF) of South Africa.
The Hobart and Ceduna antennas are operated by the University of Tasmania and this
research has been supported by AuScope Ltd., funded under the National
Collaborative Research Infrastructure Strategy (NCRIS).
The Australia Telescope Compact Array, the Parkes (Murriyang) radio telescope, and the 
Mopra telescope are all part of the Australia Telescope National Facility 
(\url{https://ror.org/05qajvd42}) which is funded by the Australian Government for 
operation as a National Facility managed by the Commonwealth Scientific and Industrial Research Organisation (CSIRO).
The YEBES40M radio telescope at Yebes Observatory is operated by the Spanish National Geographic 
Institute (IGN, Ministerio de Transportes y Movilidad Sostenible).
We are grateful to the staff of the Korea VLBI Network (KVN) who helped to operate the 
KVN array and to correlate the KVN sessions. The KVN and a high-performance computing cluster 
are facilities operated by the Korea Astronomy and Space Science Institute (KASI). 
The KVN observations and correlations are supported through the high-speed network 
connections among the KVN sites provided by the KREONET (Korea Research Environment Open NETwork), 
which is managed and operated by the KISTI (Korea Institute of Science and Technology Information).
SEJONG is managed by the National Geographic Information Institute (NGII, \url{http://ngii.go.kr}).
TIANMA65 is operated by the Shanghai Astronomical Observatory, Chinese Academy of Sciences. 
Some of this research was carried out at the Jet Propulsion Laboratory, California Institute 
of Technology, under a contract with the National Aeronautics and Space Administration (80NM0018D0004).
SSL and WYC acknowledge support from the National Research Foundation of Korea (NRF) grant 
funded by the Korea government (MIST) (2020R1A2C2009003, RS-2025-00562700).
This work has made use of data from the European Space Agency (ESA) mission
{\it Gaia} (\url{https://www.cosmos.esa.int/gaia}), processed by the {\it Gaia}
Data Processing and Analysis Consortium (DPAC,
\url{https://www.cosmos.esa.int/web/gaia/dpac/consortium}). Funding for the DPAC
has been provided by national institutions, in particular the institutions
participating in the {\it Gaia} Multilateral Agreement. Lastly we thank Dr. Nathan Secrest
and Dr. Phil Cigan for helpful comments and help in computing the VSH components and also
Dr. Ciprian Berghea and Dr. Barry Rothberg for pre-submission critical reviews.



\end{document}